# FisherMob : Un modèle bioéconomique de la mobilité des pêcheurs.

Timothée Brochier[1] et Alassane Bah[2]

[1] Institut de Recherche pour Développement (IRD), UMI 209, UMMISCO, Sorbonne Universités, Bondy, France, Campus international UCAD-IRD de Hann, Dakar, Senegal

[2] Université Cheikh Anta Diop (UCAD), Département Génie Informatique, Ecole Supérieure Polytechnique, BP 15915 Dakar, Sénégal

Timothee.Brochier@ird.fr
alassane.bah@gmail.com

**Abstract:**

Sea fishing is a highly mobile activity, favoured by the vastness of the oceans, the absence of physical boundaries and the abstraction of legislative boundaries. Understanding and anticipating this mobility is a major challenge for fisheries management issues, both at the national and international levels. "FisherMob" is a free Gama tool designed to study the effect of economic and biological factors on the dynamics of connected fisheries. It incorporate the most important processes involved in fisheries dynamics: fish abundance variability, price of the fishing effort and ex-vessel fish market price that which depends on the ratio between offer and demand. The tool uses as input a scheme of a coastal area with delimited fishing sites, fish biological parameters and fisheries parameters. It runs with a user-friendly graphic interface and generates output files that can be post-processed easily using graphic and statistical software.

**Résumé :**

La pêche maritime est un secteur d'activité où l'on observe une très forte mobilité, favorisée par l'immensité des océans, l'absence de frontières physique et l'abstraction des frontières législatives. Comprendre et anticiper ces mobilités est un défi majeur pour les questions de gestion des pêcheries, au niveau des pays comme à l'international. " FisherMob " est un outil dont l'objet est la modélisation multi-agents des interactions entre facteurs économiques et biologiques sur la distribution de l'effort de pêche et de l'impact sur la dynamique des pêcheries. Les principaux paramètres du modèle concernent la biologie des populations de poisson exploité, la distribution géographique des sites de pêche, les caractéristique de la pêcherie (coût de l'effort de pêche, capturabilité, mobilité) et enfin les caractéristiques du marché (relation de demande en fonction du prix). En fonction de ces paramètres, l'évolution de la densité de poisson, de l'effort de pêche et du prix au débarquement émergent du modèle et déterminent la dynamique de la pêcherie. L'outil possède une interface ergonomique permettant de faire varier facilement les paramètres et de visualiser les résultats des simulations en temps réel. De plus, des fichiers intégrant tous les résultats sont générés à chaque simulation permettant un post-traitement des résultats avec des outils spécialisés graphiques ou statistiques. Un premier cas d'étude concret est présenté, basé sur la pêche artisanale sénégalaise, qui a montré les potentiels de l'outil pour l'aide à la gestion des pêcheries transfrontalières. Enfin, des développements de cet outil sont suggérés.

**Mots-clés :**

Modèle individu centré, pêche, prix variable, mobilité des pêcheurs

## 1 Introduction

### 1.1 Economie, ressource et mobilité de la pêche

La dynamique des systèmes halieutiques est soumise d'une part à la dynamique de population des poissons, et d'autre part au contexte économique. Ces processus déterminent les profits réalisés par les pêcheurs, et par voie de conséquence le développement ou la récession de ce secteur d'activité. De plus, la mobilité des unités de pêche permet une grande souplesse dans la distribution spatiale de l'effort de pêche en fonction de la distribution de la ressource [1]–[3]. C'est le cas pour la pêche artisanale à l'échelle régionale [4] mais aussi pour la pêche industrielle à l'échelle des bassins océaniques [5]. Pour étudier les équilibres qui en découlent, des modèles mathématiques qui combinent la dynamique de population des poissons et la dynamique du prix en fonction de l'offre et de la demande ont été développés [6]. Par ailleurs, les modèles de dynamique spatiale des flottes de pêche ont également été développés sous différents formalismes [7]–[9]. Enfin, de récents travaux ont montré que ce sont les interactions du prix dynamique et de la mobilité des flottes de pêche qui permet de comprendre les changements observés dans certaines pêcheries [10]. En effet, en réponse à la surexploitation locale des stocks de poissons, de nombreuses pêcheries ont augmenté leur niveau de mobilité pour chercher des stocks de poissons non surexploités et maintenir leur rentabilité. Ce mécanisme d'autorégulation du système est plus important dans les zones de faible gouvernance où les pêcheries sont faiblement régulées, et les frontières maritimes peu contrôlées. Nous développons le modèle informatique « FisherMob » dans l'idée d'apporter un outil d'étude pratique de ces dynamiques en particulier pour le cas des pêcheries partagées internationales, afin d'appréhender les interactions complexes entre paramètres biologiques, économiques, et de mobilité des pêcheurs. L'outil est présenté à travers la configuration de base initialement développée pour l'étude de la dynamique de la pêcherie du mérou en Afrique de l'Ouest [10].

## 1.2 L'outil informatique

« FisherMob » est un modèle individu basé qui représente de façon explicite l'interaction entre la mobilité des pêcheurs, la dynamique de population des poissons et la dynamique du marché. La topologie des sites de pêche (leur nombre, taille, distance) et leur capacité de charge biologique (quantité maximum de poissons) sont définies par l'utilisateur en fonctions de l'état de l'art des connaissances. Le coût d'effort de pêche est également défini pour chaque zone, ainsi qu'un coût supplémentaire qui est lié à la mobilité des pêcheurs. « FisherMob » est un outil développé en Gama qui peut être téléchargé gratuitement[1], ainsi que l'environnement Gama nécessaire pour l'exécuter[2]. Le package distribué consiste en une archive compressée qui contient le code source du programme avec une configuration d'exemple.
FisherMob est paramétré via une interface graphique. Les résultats sont affichés au cours de l'exécution d'une simulation, à savoir l'évolution du nombre d'individus pêcheurs dans les différents sites de pêche, les captures, et le prix. Cet outil a initialement été développé dans le cadre de l'étude de la pêcherie du thiof au Sénégal et dans la sous-région, et a démontré son aptitude à simuler les dynamiques observées dans[1] les tendances de prix, de débarquements et d'abondances estimées [10].

## 2 Le modèle conceptuel

Dans cette section, on se propose de décrire le modèle suivant le protocole (ODD (Overview, Design concepts, Details, i.e. Aperçu Général, Concepts structurants et Détails) pour la description des modèles individus et agents ([11], [12]).

### 2.1 Objectif du modèle

L'objectif du modèle FisherMob est la mise en lumière des paramètres clefs de l'auto-organisation des pêcheries dans les zones faiblement réglementées. En particulier, le modèle vise à sensibiliser les gestionnaires sur les effets de la redistribution de l'effort de pêche sous l'effet de la mobilité économique des pêcheurs, en fonction de l'abondance de poisson, des coûts de l'effort de pêche et de la demande. Ce modèle se veut une base pour aller vers une représentation plus riche et plus réaliste des dynamiques à l'œuvre dans le monde réel, en s'ouvrant au développement par modules des différents aspect considérés.

### 2.2 Entités, variables d'état et échelles

Le modèle compte deux types d'entités qui sont les individus pêcheurs et les sites de pêche. Il y a trois variables d'états qui sont (1) la biomasse de poissons, définie pour chaque site de pêche, (2) le prix du poisson défini en fonction de la capture totale, et l'effort de pêche, qui est le nombre d'individus pêcheurs. L'espace est divisé en sites de pêche, avec des populations de poissons indépendantes les une des autres. Le prix fluctue en fonction de l'offre et de la demande.

---

[1] https://github.com/tbrochier/FisherMob
[2] www.gama.com

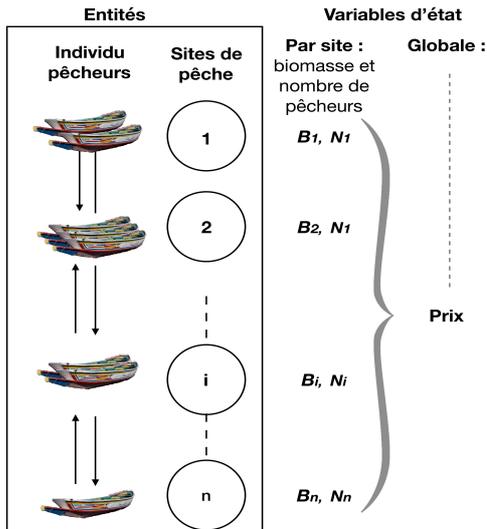

**Figure 1 :** Représentation graphique des entités et variable d'états du modèle

*Individus*. Les unités de pêche sont le seul type d'individus représentés dans le modèle. Ils partagent tous les mêmes caractéristiques, puisque l'on s'intéresse ici à la dynamique d'une pêcherie en particulier. On considère donc que chaque individu représente le même genre d'embarcation, avec le même engin de pêche et type d'équipage. Ils sont tous affectés du même niveau de capturabilité $q$, qui est défini comme la proportion de poissons présents dans un site de pêche qu'ils peuvent extraire à chaque pas de temps. Ils partagent également le même coût unitaire de base de l'effort de pêche, *i.e.* l'investissement réalisé à chaque pas de temps. Ce coût sera néanmoins corrigé par les éventuels surcoûts liés au site de pêche. Les variables d'état qui varient selon les individus sont la position dans l'espace (le site de pêche où il opère) et le profit réalisé à chaque pas de temps. Chaque pas de temps représente une campagne de pêche soit environ 15 jours dans le cas de la pêcherie de thiof au Sénégal.

*Sites de pêche*. Les sites de pêche sont des entités géographiques contiguës dans lesquelles les individus pêcheurs peuvent se déplacer. Chaque site de pêche est caractérisé par des valeurs fixes qui sont (1) la capacité de charge de la population de poisson et (2) le surcoût de l'effort de pêche dû à la mobilité requise pour s'y rendre. De plus, à chaque site de pêche est associée une biomasse de poisson, laquelle évolue à chaque pas de temps en fonction du niveau d'exploitation et de la croissance de la population. Selon les données disponibles, l'hétérogénéité des sites de pêches en terme de taille, productivité et accessibilité peut être prise en compte en affectant différentes valeurs de capacité de charge et de surcoût d'effort de pêche (voir section 4.1).

*Prix du poisson*. Le prix du poisson est déterminé en fonction de la capture totale réalisée par l'ensemble des individus pêcheurs, selon une relation empirique.

### 2.3 Vue générale des processus et ordonnancement

La représentation graphique du formalisme informatique est présentée en figure 2.

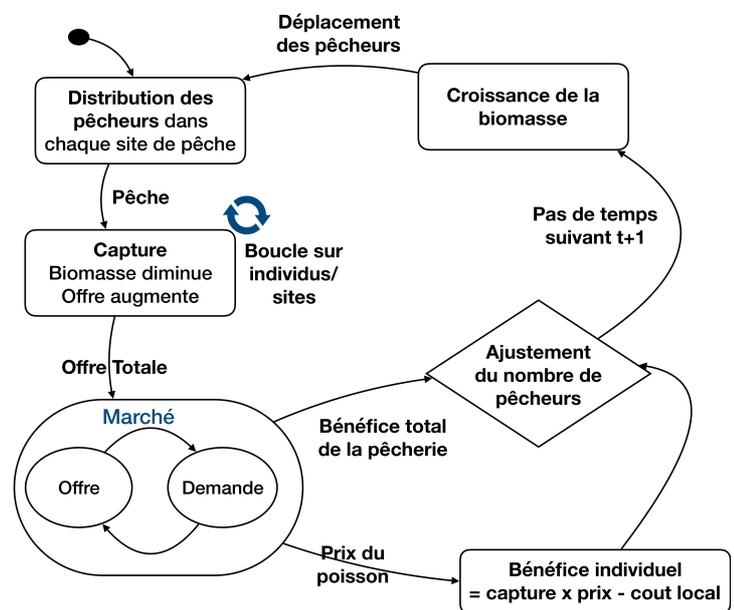

**Figure 2 :** Représentation graphique du formalisme informatique

*Calculs.* A chaque nouveau pas de temps, les processus suivant sont effectués. D'abord, la croissance des poissons dans chaque site est calculée et la biomasse actualisée (croissance logistique). Puis, l'éventuel déplacement des individus pêcheurs vers un site de pêche contiguë est déterminé selon les profits réalisés au précédant pas de temps. Ensuite, la capture de chaque individu est calculée selon la biomasse de poisson dans le site de pêche où il se trouve. La biomasse de poisson dans chaque site est réduite par l'extraction que chaque individu pêcheur présent y effectue. L'effort de pêche total est alors ajusté en fonction du bénéfice réalisé par l'ensemble de la pêcherie.

Enfin, le prix est ajusté en fonction de l'offre et la demande.

*Affichage* : A chaque pas de temps, la position des individus s'affiche sur une carte représentant la distribution spatiale des sites de pêche. De plus, sont prolongées les courbes indiquant le nombre total d'individus pêcheurs, la biomasse de poisson par site de pêche, le tonnage de poisson débarqué par les pêcheurs, le prix du poisson, la demande, et le bénéfice total de la pêcherie.

*Enregistrement* : À chaque pas de temps, une nouvelle ligne est écrite dans un fichier de sauvegarde des résultats de la simulation au format « csv ».

## 2.4 Conceptualisation

*Principes de base*. La dynamique de population des poissons est représentée par une croissance logistique de la biomasse accessible à la pêche. Le prélèvement de biomasse des individus pêcheurs est calculé comme une fraction de la biomasse. Il est considéré que la mobilité des pêcheurs vers des sites de pêche distant induit un coût supplémentaire. La dynamique du prix se base sur une relation entre le prix et la demande, en considérant l'existence d'une demande maximum si le prix tend vers zéro. Enfin, on considère que les pêcheurs ont la possibilité de rapidement se tourner vers d'autres activités lorsqu'ils n'effectuent plus de profit en ciblant une espèce particulière (et donc de sortir de la pêcherie spécifique considérée dans le modèle). Cela peut correspondre à un changement de stratégie ou de technique de pêche comme cela est couramment observé dans la pêche artisanale sénégalaise [13].

*Émergence*. L'équilibre entre la biomasse de poisson, le nombre de pêcheurs et le prix émerge en fonction des paramètres biologiques, économiques et des caractéristiques des pêcheurs. L'étude du modèle permet de montrer à quelles conditions l'auto-organisation du système peut mener à l'émergence d'un équilibre permettant le maintien durable de l'activité de pêche.

*Adaptation*. Les individus pêcheurs adaptent leur position (site de pêche) en fonction du profit qu'ils ont réalisé au pas de temps précédant. Si le profit était négatif, ils se déplacent aléatoirement vers l'un des sites adjacents. Par ailleurs, si le bénéfice total est inférieur à zéro, le nombre total d'individus pêcheurs est réduit. Si au contraire le bénéfice est positif, le nombre d'individus pêcheurs est augmenté. L'augmentation ou la diminution de l'effectif des pêcheurs se fait par ajout/retrait d'individus aléatoirement sur les sites de pêche.

*Perception/Collectivité*. La perception des individus pêcheurs de la profitabilité d'un site de pêche se fait selon les bénéfices qui seraient réalisés si la pêche était menée, qui combinent la capture, le prix (valeur débarquée), et le coût de l'effort de pêche. On considère que ces informations peuvent être estimées par les pêcheurs grâce à la communication qu'il existe entre eux. Par ailleurs, nous avons considéré une perception globale de la rentabilité de la pêcherie, pour traduire le fait que l'unité de pêche se trouvant sur différents sites peuvent parfois appartenir à un même armateur ou une même famille pour la pêche artisanale. Cette perception globale de l'effort de pêche détermine l'augmentation ou la diminution de l'effort total (voir *Adaptation*).

*Interactions*. Les individus pêcheurs interagissent d'une part via la compétition directe pour l'exploitation de la ressource, et d'autre part via l'effet sur le prix global de la mise sur le marché de leur débarquement. Ainsi, des pêcheurs opérant dans des sites distants participent à satisfaire la même demande.

*Stochastique*. Le déplacement des individus vers l'un ou l'autre des sites de pêche adjacents se fait de manière aléatoire.

*Observation*. L'effort de pêche, la capture, et le prix sont les principales variables à observer pour comprendre les évolutions du modèle. La distribution spatiale de l'effort de pêche est calculée en réalisant la somme du nombre d'individus pêcheurs dans chaque site de pêche. On peut ainsi agréger l'effort de pêche par pays, si les sites de pêche considérés couvrent plusieurs pays. De même la capture totale des individus pêcheurs et les biomasses de poissons peuvent être agrégées par pays. Le prix et la demande sont des variables non spatialisées dans le modèle. Enfin, le bénéfice total de la pêcherie est la somme des bénéfices réalisés par tous les individus.

## 2.5 Initialisation

On démarre la simulation avec un faible nombre de pêcheurs distribués sur une partie des sites de pêche pour lesquels les frais de mobilité sont nuls. La biomasse de poissons initiale dans chaque site de pêche est vierge, c'est à dire égale à la capacité de charge. Le prix initial est faible, correspondant aux plus anciennes données disponibles. Très rapidement la dynamique du modèle opère et converge soit vers un équilibre soit vers une extinction rapide de la pêcherie. On montre qu'il existe une large gamme de conditions initiales sur le prix qui permettent le développement de la pêcherie. Dès lors que la pêcherie connaît un développement initial, les conditions initiales n'influencent plus la situation à long terme.

## 2.6 Données d'entrée du modèle (*inputs*).

Le modèle n'utilise pas de données d'entrée pour représenter des évolutions temporelles des processus. Au contraire, l'objectif du modèle vise à l'émergence de ces évolutions temporelles par la simulation des processus sous-jacents (voir *émergence*). Ainsi, les données de séries temporelles sur la dynamique des pêcheries qui auront pu être rassemblées ne constituent pas une entrée du modèle. Ces séries temporelles seront plutôt utilisées comme source de questionnement au modèle, lequel proposera une dynamique sous-jacente.

## 2.7 Sous-modèles.

***Croissance de la biomasse des poissons***. La croissance de la biomasse $B_i$ au site de pêche $i$ est modélisée par une loi logistique, comme proposé par Schaefer [14].

$$B_{i\,t+1} = B_{i\,t} + \left(1 - \frac{B_{i\,t}}{K}\right) r B_{i\,t} \quad (1)$$

Où $r$ est le coefficient de croissance de la population et $K$ la capacité de charge du milieu (site de pêche).
Dans cette version du modèle, nous avons considéré des poissons sédentaires ; il n'y a donc pas de terme de migration. Par contre, la biomasse de poisson subit les prélèvements de la pêche réalisée par chaque individu pêcheur.

***Prélèvement de la biomasse de poissons par les pêcheurs***. C'est l'action de pêcher, elle est donnée par la séquence suivante, appliquée pour chaque pêcheur dans chaque site de pêche au moment de la boucle sur les individus. On enregistre la capture, pour chaque individu pêcheur présent au site $i$, et on actualise la biomasse locale de poissons

$$Capture = qB_i \quad (2)$$
$$B_{i\,actualisé} = B_i - qB_i \quad (3)$$

Où $q$ est la capturabilité et $B_i$ la biomasse au site $i$.

***Bénéfice réalisé par les pêcheurs***. Le bénéfice réalisé lors de chaque sortie de pêche pour chaque individu situé dans le site de pêche $i$ est calculé de la manière suivante :

$$bénéfice = pqB_i - c_i \quad (4)$$

Où $q$ est la capturabilité, $p$ le prix, $B_i$ la biomasse au site $i$ et $c_i$ le coût de la sortie de pêche dans la zone $i$, définit comme suit :

$$c_i = c + \delta \quad (5)$$

Avec c le coût constant d'une sortie pêche et δ le surcoût dû à la mobilité dans ce site de pêche.

***Déplacement des pêcheurs***. On considère que les individus pêcheurs ne se déplacent que s'ils ne réalisent plus de bénéfice dans la zone où ils se trouvent. Du fait de la communication entre les pêcheurs, nous posons l'hypothèse qu'un individu peut avoir une idée du bénéfice potentiel avant de sortir en mer.
Pour implémenter cette hypothèse, au début du pas de temps, on calcule le bénéfice qui serait réalisé en cas de pêche dans le site actuel. Si le bénéfice estimé est positif, le pêcheur reste sur place. S'il est négatif, le pêcheur se déplace aléatoirement dans l'un des 2 sites voisins, et l'opération est répétée. Si au bout du troisième déplacement le bénéfice estimé est toujours négatif, l'individu quitte la pêcherie considérée (voir section 2.4, *principes de base*).

***Ajustement de l'effort de pêche total***. D'abord, le bénéfice total de la pêcherie est calculé au cours de la boucle sur les individus en cumulant les bénéfices réalisés par chaque pêcheur. Après la boucle sur l'ensemble des individus, le nombre total d'individus est calculé en fonction du bénéfice total réalisé par la pêcherie de la manière suivante :

$$N(t+1) = \phi \frac{\text{Benefice Total}}{c} \quad (6)$$

**c** = coût unitaire de l'effort de pêche ; Φ = taux de réinvestissement ou désinvestissement.

L'idée est que l'investissement total dans l'effort de pêche général au temps t+1 correspondra aux bénéfices réalisés l'année précédente.

***Ajustement du prix.*** Suivant la formulation de [6], la demande est calculée, en fonction du prix au temps t selon la formule suivante :

$$Demande = \frac{A}{1+\beta p_t} \quad (7)$$

Puis, le prix au temps t+1 est donné par la formule suivante :

$$p_{t+1} = p_t + \varphi(Demande - Capture) \quad (8)$$

Où « capture » est la capture totale réalisée au pas de temps t par l'ensemble des pêcheurs.

## 3 Cas d'étude : la pêche du thiof en Afrique de l'ouest

### 3.1 Configuration

A titre d'illustration, nous présentons ici un exemple d'application à l'étude de la pêcherie artisanale sénégalaise du thiof (mérou), en Afrique de l'Ouest. Nous avons considéré 20 sites de pêche représentant chacun 100km de côte linéaire entre la Guinée et la Mauritanie. La figure 3 présente une vue du découpage des sites de pêche. Les poissons sont répartis uniformément dans l'espace, chaque site de pêche étant affecté d'une même capacité de charge $K$. Notons que le modèle considère une pêcherie mono-spécifique. La part des frais de mobilité dans le coût de l'effort de pêche déployé augmente lorsque l'on s'éloigne du Sénégal. Les paramètres de la configuration sont présentés dans le tableau 1.

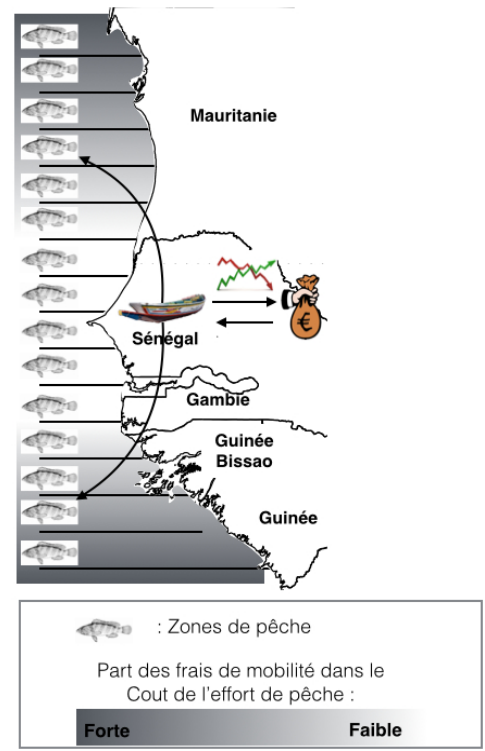

**Figure 3 :** Les sites de pêches du thiof accessible à la pêche artisanale sénégalaise.

Initialement, les individus pêcheurs sont distribués sur les sites de pêche localisés au Sénégal. De l'année 1 à 37, les mobilités sont limitées à l'intérieur de la zone Sénégal, puis à partir de l'année 38 on active le bouton « migrations autorisée » qui permet à l'effort de pêche de se distribuer dans les pays voisins (Mauritanie et Guinée, cf fig. 1), avec des coûts additionnels.

A chaque pas de temps, sont enregistrées la capture totale, la demande, la biomasse en Guinée, en Mauritanie, au Sénégal, Le nombre de pêcheurs au Sénégal, en Mauritanie, en guinée, le bénéfice total, et enfin le prix (voir section 2.4 *Observation*).

**Tableau 1:** Liste des paramètres du modèle, avec valeurs indicatives pour la configuration d'exemple (« pêcherie artisanale sénégalaise de thiof »).

| \multicolumn{3}{c}{*Paramètres de Dynamique de Population des Poissons*} |||
|---|---|---|
| Symbole | Description | Valeur et unité |
| r | Taux de croissance de biomasse | 0.2 ans$^{-1}$ |
| K | Capacité de charge de l'environnement | 10 000 tonnes |
| L | Nombre de lieux de pêche | 20 |
| F | Nombre de lieux de pêche accessibles | 1-20 |
| δ | Facteur de surcoût dû à la mobilité | 10 |
| q | Capturabilité | 0.01 |
| $C_{local}$ | Coût d'une sortie de pêche | 200 000 XOF |
| Φ | Taux de réinvestissement | 0.01 Effort / XOF |
| p | Prix (si constant) | 100 000 XOF / tonne |
| A | Demande max | 100 000 tonnes |
| β | Sensibilité du prix | 0.01 10$^{-6}$ XOF |

### 3.2 Résultats

Les résultats des simulations qui montrent l'évolution des variables d'état du système sont présentés en figure 2.

Avant l'autorisation des migrations, le modèle montre une baisse continue du stock de poisson au Sénégal, concomitante avec une hausse du prix du poisson et une augmentation de l'effort de pêche ; en somme une situation typique de surexploitation.

Immédiatement après l'autorisation des migrations, le nombre de sorties de pêche augmente significativement (~de 50%) et brutalement tant au Sénégal qu'en Mauritanie et en Guinée (Fig. 1a). Les biomasses de poisson en Mauritanie et en Guinée connaissent une chute significative (~de -70%) et brutale ; au Sénégal en revanche la tendance s'inverse et entame une légère hausse (Fig. 1b). La capture totale augmente aussi brutalement, et la demande repart à la hausse mais reste inférieure à l'offre (Fig. 1c). Enfin, le profit total de la pêcherie connaît une forte et courte augmentation au début des migrations, puis se stabilise rapidement à une valeur légèrement supérieure à la situation avant migration. Le prix du poisson, qui était à la hausse, repart à la baisse dès le début des migrations (Fig. 1d).

Finalement, il apparaît que l'autorisation de la migration dans le modèle a permis de passer d'une dynamique de surexploitation au Sénégal à une dynamique que l'on pourrait qualifier de « pêche durable », avec une stabilité des stocks et des prix. La brutalité des variations de l'effort de pêche, des débarquements et de la biomasse est certainement due au caractère abrupt du scénario simulé, où les migrations hors Sénégal commencent d'un coup. Dans un scénario plus réaliste, l'effort de pêche à l'international augmenterait plus graduellement car limité par le contrôle aux frontières. Remarquons tout de même que des variations brutales ont effectivement été observées au Sénégal pour d'autres pêcheries, avec des périodes d'offre qui dépassent la demande pour certaines espèces comme par exemple le poulpe [15]. De telles variations brutales, sont également reproduites de manière saisonnière dans le modèle de pêche artisanale de Le Fur [16], et décrites comme les « pulsations » du système.

Nous renvoyons le lecteur à [10] pour l'analyse détaillée de ces résultats au vu des observations disponibles au Sénégal et dans la sous-région.

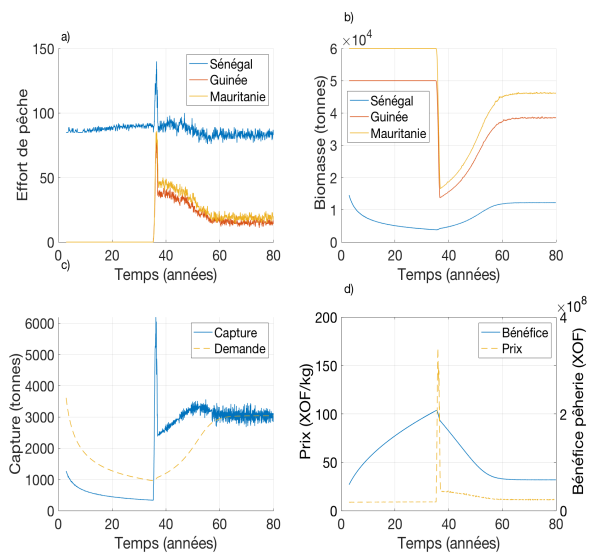

**Figure 4 :** Résultats de simulation de la pêcherie artisanale sénégalaise de thiof

## 4 Conclusion et perspectives

Les bases d'un modèle individu de la mobilité de la pêche artisanale ont été posées. Celui-ci a permis de reproduire les grandes lignes des dynamiques observées dans un cas réel [10]. La description selon le protocole ODD permet d'en clarifier les choix, et fournit une documentation détaillée. L'affinement des sous-modèles pourra permettre de faire évoluer l'outil pour reproduire et expliquer des processus de plus en plus détaillés de la réalité, afin d'en faire un véritable outil d'aide à la gestion des pêches dans les zones faiblement réglementées. L'exploration de la sensibilité du modèle aux paramètres choisis grâce à des algorithmes adaptés (e.g. openmole [17]) permettront d'assurer la généralisation et la robustesse des résultats obtenus, et de calibrer les paramètres du modèle aux observations.

### 4.1 Vers un environnement réaliste

La lecture des fichiers de sortie des modèles hydrodynamiques et biogéochimiques disponibles pour la sous-région (e.g. [18]) permettra de rendre plus réaliste la description des sites de pêche et donc d'intégrer divers paramètres qui influenceront la capacité de charge des poissons selon les sites de pêche (e.g. trait de côte, bathymétrie, courants) et même de prendre en compte les aspects de migration saisonnière de la ressourc [19], et de connectivité larvaire des stocks de poissons [20]). Les fichiers de sortie de ces modèles sont disponibles au format de donnée netcdf.

### 4.2 Questions de recherche ouvertes

La structure présente du modèle permet de mettre en évidence une auto-organisation de la distribution de l'effort de pêche sous l'impulsion des facteurs de surexploitation et de variabilité du prix. Si ce mécanisme existe probablement dans la réalité, il est clair qu'il se superpose à d'autres facteurs, telle que la variabilité environnementale. Pour comprendre et anticiper la réorganisation continue de la pêche artisanale, il sera intéressant de tenter de reproduire des schémas migratoires à l'échelle saisonnière comme observée au Sénégal. L'identification des effets relatifs de l'environnement et de l'économie sur la mobilité des pêcheurs devra reposer sur une approche participative au cours de laquelle les hypothèses du modèle seront appelées à évoluer, comme par exemple l'hypothèse d'homogénéité des individus pêcheurs et l'existence de coûts liés à la reconversion des pêcheurs [21].